# High-pressure phases of Weyl semimetals NbP, NbAs, TaP and TaAs

Zhaopeng Guo, Pengchao Lu, Tong Chen, Juefei Wu, and Jian Sun*, Dingyu Xing

*National Laboratory of Solid State Microstructures,*

*School of Physics and Collaborative Innovation Center of Advanced Microstructures,*

*Nanjing University, Nanjing, 210093, P. R. China*

**ABSTRACT**

The high pressure phase diagrams of TaAs family (NbP, NbAs, TaP and TaAs) are explored systematically using crystal structure search together with first-principles calculations. Our calculations show that NbAs and TaAs have similar phase diagrams and possess the same structural phase transition sequence of $I4_1md$ → P-6m2 → $P2_1/c$ → Pm-3m, with a bit different transition pressures. The phase transition sequence of NbP and TaP are somewhat different from that of NbAs and TaAs. In which new structures emerge, for instance, the *Cmcm* structure in NbP and the *Pmmn* structure in TaP. It is found interestingly that in the electronic structure of the high-pressure phase *P-6m2*-NbAs, there are Weyl points and triple degenerate points coexisted, similar to the high-pressure *P-6m2*-TaAs.

* Correspondence should be addressed to J.S. (Email: jiansun@nju.edu.cn)

## I. INTRODUCTION

The topological semimetals (TSMs) with non-trivial band structure topology have been recently studied both theoretically and experimentally [1-3]. Depending on the shape and degeneracy of band crossings around the Fermi level, the TSM can be mainly classified into the Dirac semimetal (DSM) [4-6], the Weyl semimetal (WSM) [7-18], and the topological nodal line semimetal (NLSM) [19-23], etc. Besides, there are three-, six-, and eight-band crossings, as mentioned in the work of Bradlyn *et al.*[24]. Recently, the first-principle calculations on WC-type ZrTe [25,26] and $\theta$-phase TaN [27] showed that those materials contain triple degenerate points (TDPT) [28-33]. This type of fermions was recently been proved with angle-resolved photoemission spectroscopy (ARPES) experiment in MoP [34].

The WSM can be realized when the system breaks either time reversal symmetry or spatial inversion symmetry. The WSM has special features in surface state and electric transports, such as discontinuous Fermi arc [7-9] and negative magnetoresistance [10-14] due to the Adler-Bell-Jackiw anomaly. The idea of WSM was theoretically proposed by Wan *et al.* in 2011 [7]. The earliest predicted WSM material is pyrochlore iridate, which is somehow difficult for sample growth and experimentally verification due to its magnetism. Recently, noncentrosymmetric transition-metal monophosphides were proposed [8] and experimental confirmed using ARPES [15-18] and negative magnetoresistance measurements [12,35,36]. The WSM can also be classified into Type-I and Type-II. The Type-I WSMs such as TaAs [8] respect Lorentz symmetry, but type-II WSMs such as WTe2 [37] and MoTe2 [38] do not. Different from the Type-I WSM, the main magneto-transport feature of the Type-II WSM is the anisotropic chiral anomaly [37]. The TDPT mentioned above also has the direction dependence in magnetoresistance named as “helical anomaly” [25,27].

High pressure is a clean way to modify crystal structure without intruding impurities and defects. Variation of crystal structure usually leads to changes in electronic structures and topological states. Thus, it is interesting to investigate high pressure phase transitions of materials, especially for topological insulators and topological semimetals. For instance, pressure-induced superconductivity of topological insulator such as $Bi_2Te_3$ [39] and $Bi_2Se_3$ [40] was observed. And pressure-induced superconductivity in WSM was observed as well, such as $WTe_2$ [41-43] and TaP [44]. The high pressure phase of TaAs is a new topological WSM [45]. While several high pressure phases of TaAs family were studied discretely [44-47], a systematic study on the high pressure phase diagrams for the whole family (NbP, NbAs, TaP, and TaAs) is still lacking.

In this work, we study the high pressure phases of the whole TaAs family, including NbP, NbAs, TaP, and TaAs, by first-principles calculations and crystal structure search method. Our calculations show that NbAs and TaAs have similar phase diagrams, while NbP and TaP are slightly different. Moreover, both NbP and TaP are harder against compression. It is interestingly found that in the high pressure phase of *P-6m2* NbAs, there are Weyl points and TDPTs coexisted, analogous to the high pressure phase of *P-6m2* TaAs [45] as well as ZrTe [25,26].

## II. COMPUTATIONAL METHOD

The ab-initio random structure searching (AIRSS) method [48,49] was used for high pressure crystal structure predictions. Structural optimizations and enthalpy calculations were performed using projector augmented wave (PAW) implemented in the Vienna ab initio simulation package (VASP) [50], with Perdew-Burke-Ernzerhof (PBE) [51] generalized gradient approximation (GGA) exchange-correlation density functional. The cutoff energy of plane wave basis was set to 400 eV and the Brillouin zones were sampled with Monkhorst-Pack method in *k*-mesh spacing of $0.03Å^{-1}$. Electronic structures calculations were performed using the full-potential linearized augmented plane-wave (FP-LAPW) method [52,53] implemented in the WIEN2k package [54]. We used a 1000 k-point mesh for Brillouin zone (BZ) sampling and -7 for the plane wave cut-off parameter $R_{MT}K_{max}$ in the calculation, where the $R_{MT}$ is the minimum muffin-tin radius and $K_{max}$ is the plane-wave vector cut-off parameter. Spin-orbit coupling (SOC) was taken into account by a second-variation method [55]. The maximally localized Wannier functions (WLWF) were constructed by wannier90 code [56], in the subspace spanned by the d orbitals of Nb atoms and p orbitals of As atoms. The projected surface states were calculated using surface Green's function in the semi-infinite system [57,58].

## III. RESULTS AND DISCUSSIONS

### A. Phase diagram and Crystal structures

To study the high-pressure phase diagram of NbP, NbAs, TaP, and TaAs (known as TaAs family) compounds, energetically favorable structures of TaAs family under pressure up to 100GPa were predicted by AIRSS method up to 100GPa. At the ambient condition, the compounds of TaAs family share the same phase of $I4_1md$ (Space group No.109) which is well known as the WSM phase. Under compression, it is expected that the ambient phase will transform to high-pressure phases. Combining random search method with first-principles calculations, we obtain several candidates of energetically stable structures and calculate their enthalpies under pressure. The enthalpies per formula relative to the ambient phase are shown by the enthalpy-pressure curves in Fig 1. The lowest-enthalpy phases of the TaAs family change with pressure, as shown in Fig 2(a), and the crystal structures of the thermodynamically stable phases are shown in Fig 2(b). Note that the structures sharing the same space group are quite similar, only the atomic positions are a little bit different. Such as the ambient phase $I4_1md$ (No.109) and high pressure phase $P2_1/c$ (No.14), which appear in the four compounds in the TaAs family. The detailed structure information of calculated and experimental structures of the TaAs family is given in Table 1.

It is interesting to see that NbAs undergoes a complicated phase transitions with increasing pressure. When compressing the NbAs crystal, its structure phase transition occurs at 31GPa, from the ambient-pressure face-centered tetragonal $I4_1md$-NbAs phase to a hexagonal *P-6m2*-NbAs (No.187) phase. Further compression leads to another structure transition at 31GPa from *P-6m2*-NbAs to *P21/c*-NbAs phase. The $P2_1/c$-NbAs, which has four formula units in the unit cell, is stable between 31GPa and 65GPa, and the cubic *Pm-3m*-NbAs is stable above 65GPa. The coordination number of the Nb atom is six in the ambient-condition phase $I4_1md$-NbAs, and its coordination numbers are six, seven, and eight, respectively, in the *P-6m2* phase, $P2_1/c$ phases, and *Pm-3m* Phase. As a result, the coordination number of the Nb atom increases under compression, which can be understood by the fact that the crystalline structure becomes denser and denser with increasing pressure.

The phase diagram of TaAs is similar to that of NbAs. The ambient phase $I4_1md$-TaAs is stable below 13GPa, and a structural phase transition occurs at 13GPa from $I4_1md$-TaAs to *P-6m2*-TaAs. Resembling to NbAs, the $P2_1/c$ phase of TaAs becomes the most stable phase above 34GPa. The similarity in phase transitions between NbAs and TaAs suggests that elements in the same family generally have similar properties. Previous studies on the high pressure phases of TaAs [45,46] more or less gave the same structure sequence.

The high-pressure phase transitions of NbP and TaP are quite different from those of NbAs and TaAs. Upon compression, we find that a phase transition from $I4_1md$-NbP to *Cmcm*-NbP (Space group No.63) occurs at 62GPa. The coordination number of the Nb atom in *Cmcm*-NbP is seven, which is higher than that in ambient phase $I4_1md$. The structure phase transitions of TaP are also different from NbP. The ambient phase $I4_1md$-TaP is stable from 0 to 69GPa, and *Pmmn*-TaP phase is the most stable structure from 69GPa to 90GPa. Above 90GPa, the $P2_1/c$-TaP is stable. In *Pmmn*-TaP (Space group No.59) phase, the coordination numbers of the Ta atoms are five for two Ta atoms at Wyckoff position 2a (0, 0, 0.0826) and seven for other two Ta atoms at position 2b (0, 0, 0.3326). The coordination number of the Ta atom in $P2_1/c$-TaP is seven, which is same as that in $P2_1/c$-NbAs. The critical pressures for the phase transitions in NbP and TaP are much higher than those in NbAs and TaAs.

## B. Electronic structures

The band structures and Fermi surfaces of ambient phase of TaAs family have been calculated. For the ambient phase $I4_1md$, if the SOC is absent, there are several nodal rings of bands crossing in the mirror planes. When taking the SOC into account, the nodal rings will be reduced into three pairs of Weyl points. The calculated electronic structures of several energy favorite structures for high-pressure phase are shown in Figs. 3 and 4, where the fat band plot method has been applied to mark the character of different orbital's contribution. The color red band means the contribution from the transition metal atom's d orbital, the color green band means the contribution from the phosphorus family atom's p orbital, and the color yellow band means the contribution from both of them.

The band structure and Fermi surfaces of *Cmcm*-NbP at 70GPa are shown in Fig 3(a). The band structure calculation shows that *Cmcm*-NbP is a metal. Its Fermi surface has many pockets which are not located around high-symmetry points, although the band structure is quite clean around the Fermi level along the high-symmetry line. With compression, the Weyl semimetal phase $I4_1md$-NbP will turn into the metal phase *Cmcm*-NbP. The electronic structures of high pressure phases of NbAs are shown in Fig 3(b) (*P-6m2*-NbAs at 35GPa), Fig 3(c) ($P2_1/c$-NbAs at 40GPa), and Fig 3(d) (*Pm-3m*-NbAs at 70GPa), respectively.

In the *P-6m2*-NbAs phase, the $\Gamma - A$ and $K - \Gamma$ band crossings are contributed by the d orbitals of the Nb atoms, and the pockets of around M point are mainly contributed by the p orbitals of As atoms. The *P-6m2*-NbAs is a compound with both triple points and Weyl points coexisted, whose topological properties will be discussed in the next section. The Fermi surface of *P-6m2* NbAs contains three kinds of pockets: the hole-pocket on the $\Gamma MK$ plane which encloses Weyl points around the K point, the electronic-pocket and the ball-like hole-pocket around the A point. We show electronic structures of high pressure phases of TaP and TaAs in Fig 4(a) for *Pmmn*-TaP at 70GPa, Fig 4(b) for $P2_1/c$-TaP at 90GPa, Fig 4(c) for *P-6m2*-TaAs at 20GPa, and Fig 4(d) for $P2_1/c$-TaAs at 40GPa. It is found that all of the high pressure phases of TaP, TaAs, and NbAs are metallic.

### C. Coexistence of Weyl points and triple points in the *P-6m2* NbAs

The electronic band structure of *P-6m2*-NbAs is quite clean near the Fermi level and the Fermi pockets are rather small. Further calculations show that there are six pairs of Weyl points in the first Brillouin zone, similar to the high pressure phase of TaAs [45] as well as the *P-6m2* ZrTe [25,26]. Each pair of Weyl points is located on the $\Gamma KA$ plane near the K point, and two paired Weyl points are related to each other by mirror plane $\Gamma KM$. The pairs of Weyl points are related to each other by the $C_3$ rotation operation and the $\Gamma MA$ mirror symmetry operation. As a result, all the Weyl points have the same energy, which is about 170 meV above the Fermi level in *P-6m2* NbAs. The band structures around the Weyl points are shown in Figs. 5(a) and 5(b). Along the $\Gamma - A$ high symmetry line, there are four TDPTs protected by the $C_{3v}$ symmetry.

To further study the topological properties of *P-6m2*-NbAs, we have performed calculations with the wannier90 code [56] together with the Wannier_tools [59]. The $\Gamma MK$ plane ($k_z$=0) and $ALH$ plane ($k_z=\pi$) can be viewed as two-dimension insulators with time reversal symmetry, so that the topology of this two-dimension system can be classified by the $Z_2$ topological invariant. The calculated $Z_2$ topological invariants using Wilson loop method [60] are shown in Figs. 5(c) and 5(d). The $Z_2$ topological invariants of $k_z$=0 and $k_z=\pi$ planes are both odd, and so both of the $\Gamma MK$ plane and $ALH$ plane are topological non-trivial.

Figure 6(a) shows the first BZ as well as the projected (100) surface of *P-6m2*-NbAs. Based on the tight-binding model, the surface states of the As-terminated (100) surface are shown in Fig 6(b), and the Fermi arc connecting the projections of Weyl points are shown in Figs 6(c) and 6(d). Near the $\overline{M}$ point, there are projections of Weyl points ($W_S$), which are marked with white circles, as shown in Fig 6(c). It is noted that this pair of projected Weyl points are inside projected Fermi pocket of the bulk state near the $\bar{M}$ point. The other projected Weyl points ($W_D$) are not connected by bulk projected Fermi surface. The detailed Fermi surface in the white frame in Fig 6 (c) is shown in Fig 6(d), where the white circle marks the location of projected Weyl points. It is noted that two Weyl points with the same chirality are projected to the same position, so that there are two Fermi arcs marked by SS1 and SS2. One Fermi arcs (SS2) begins from one Weyl point and ends at the projected bulk state; the other Fermi arc (Surface State: SS1) begins from one Weyl point and ends at another Weyl point on the other side. Similar to *P-6m2*-NbAs, the high pressure phase *P-6m2*-TaAs is also a compound with the Weyl points and triple points coexisted [45].

## IV. Conclusion

In this work, we systematically study the high pressure phase diagram of the TaAs family using AIRSS and ab initio calculations. From our calculations, it is found that one high pressure phase of NbP has the *Cmcm* structure above 60GPa, and two high pressure phases of TaP have the *Pmmn* and $P2_1/c$ structures. The high-pressure phase diagrams of NbAs and TaAs are quite complicated. For example, there are three high pressure structures of NbAs: the hexagonal *P-6m2* phase (stable for 35-40GPa), monoclinic $P2_1/c$ phase (stable for 40-70GPa), and cubic *Pm-3m* phase (stable above 70GPa). All of the high pressure phases of the TaAs family are metallic. Very interestingly, the electronic structure of *P-6m2*-NbAs is found to contain the Weyl points and triple degenerate points coexisted. The direction dependence of negative magnetoresistance will be a strong signature of the triple degenerate points in *P-6m2*-NbAs, which can be measured by future experiments.

## ACKNOWLEDGMENTS

We are grateful for the financial support from the National Key R&D program of China (Grant No: 2016YFA0300404), the National Key Projects for Basic Research in China (Grant No.2015CB921202), the National Natural Science Foundation of China (Grant Nos: 11574133 and 51372112), the NSF Jiangsu province (No. BK20150012), the Science Challenge Project (No. TZ2016001), the Fundamental Research Funds for the Central Universities and Special Program for Applied Research on Super Computation of the NSFC-Guangdong Joint Fund (the 2nd phase). Part of the calculations were performed on the supercomputer in the High Performance Computing Center of Nanjing University and "Tianhe-2" in the NSCC-Guangzhou.

# Figures

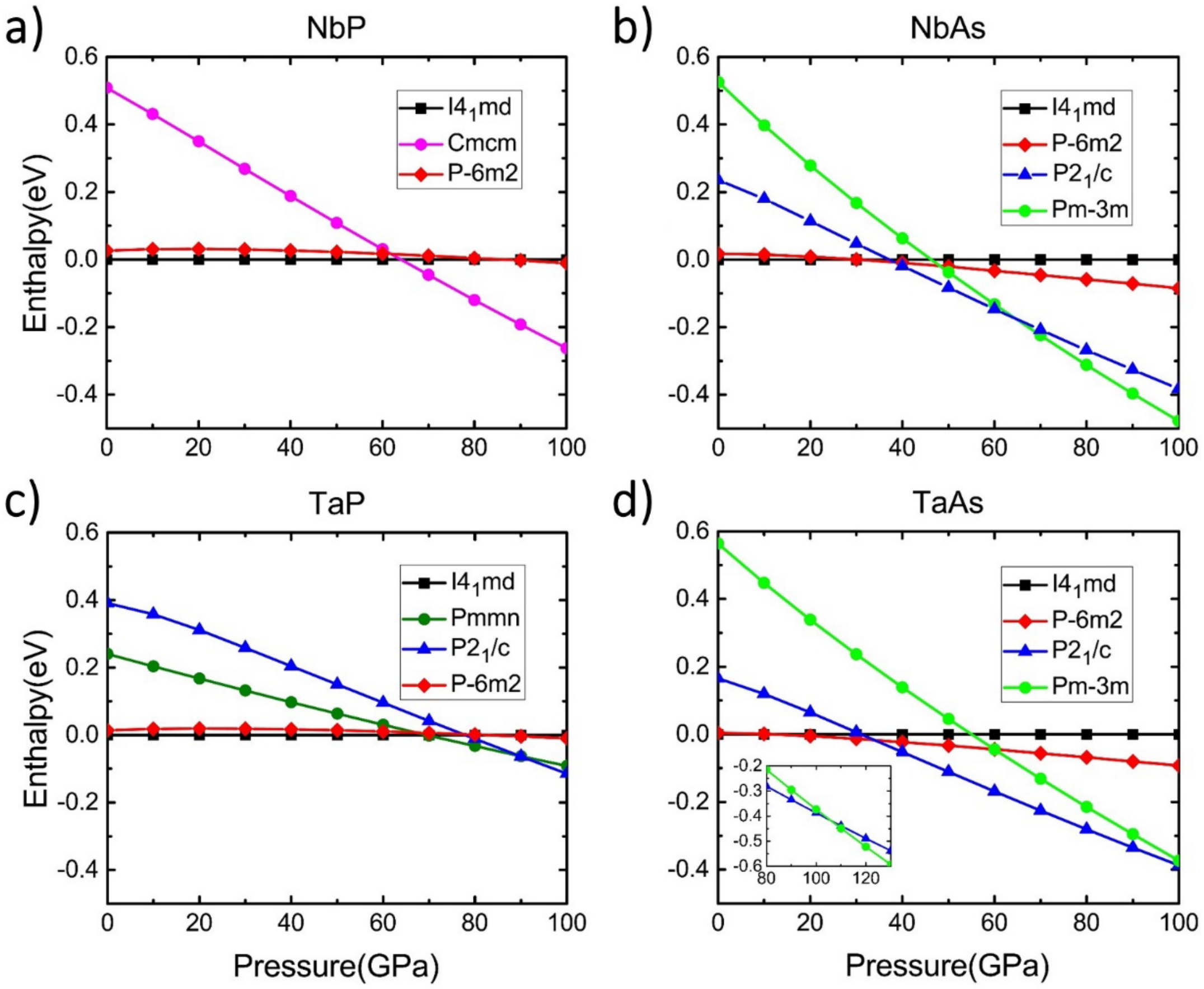

Fig 1 Enthalpy-pressure curves of NbP(a), NbAs(b), TaP(c), and TaAs(d), respectively. The enthalpies are calculated from 0GPa to 100GPa. All the enthalpies are normalized to that of per formula of NbP, NbAs, TaP and TaAs.

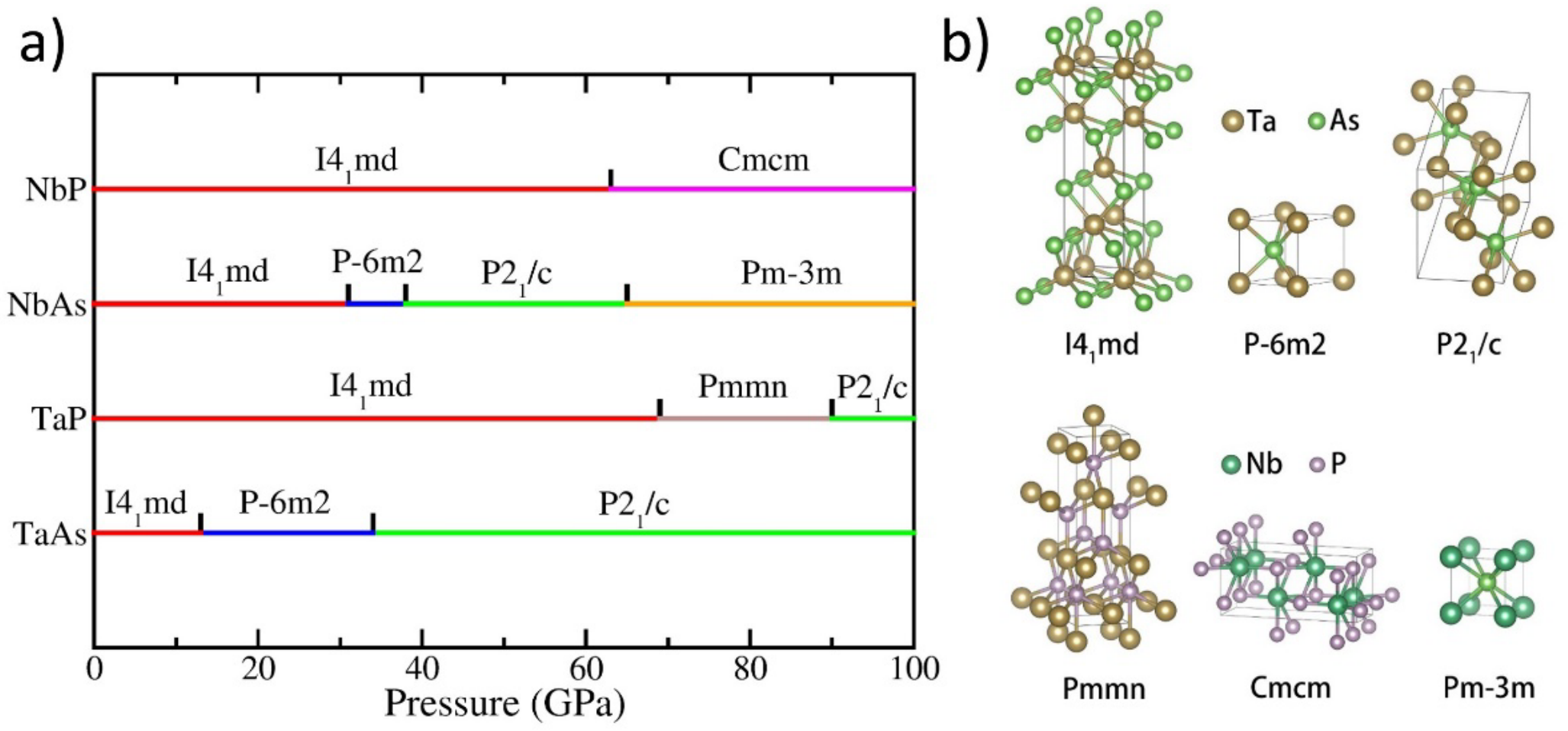

Fig 2, Phase diagram of TaAs family (a) and candidate structures (b). The stable region of ambient phases are marked with red line and blue lines represent *P-6m2* phase. It is clear that high pressure phase transition is pretty hard for NbP and TaP. In Fig (b), we show several candidate structures including the ambient structure and high pressure phases of TaAs, *Pmmn*-TaP, *Cmcm*-NbP, and *Pm-3m*-NbAs.

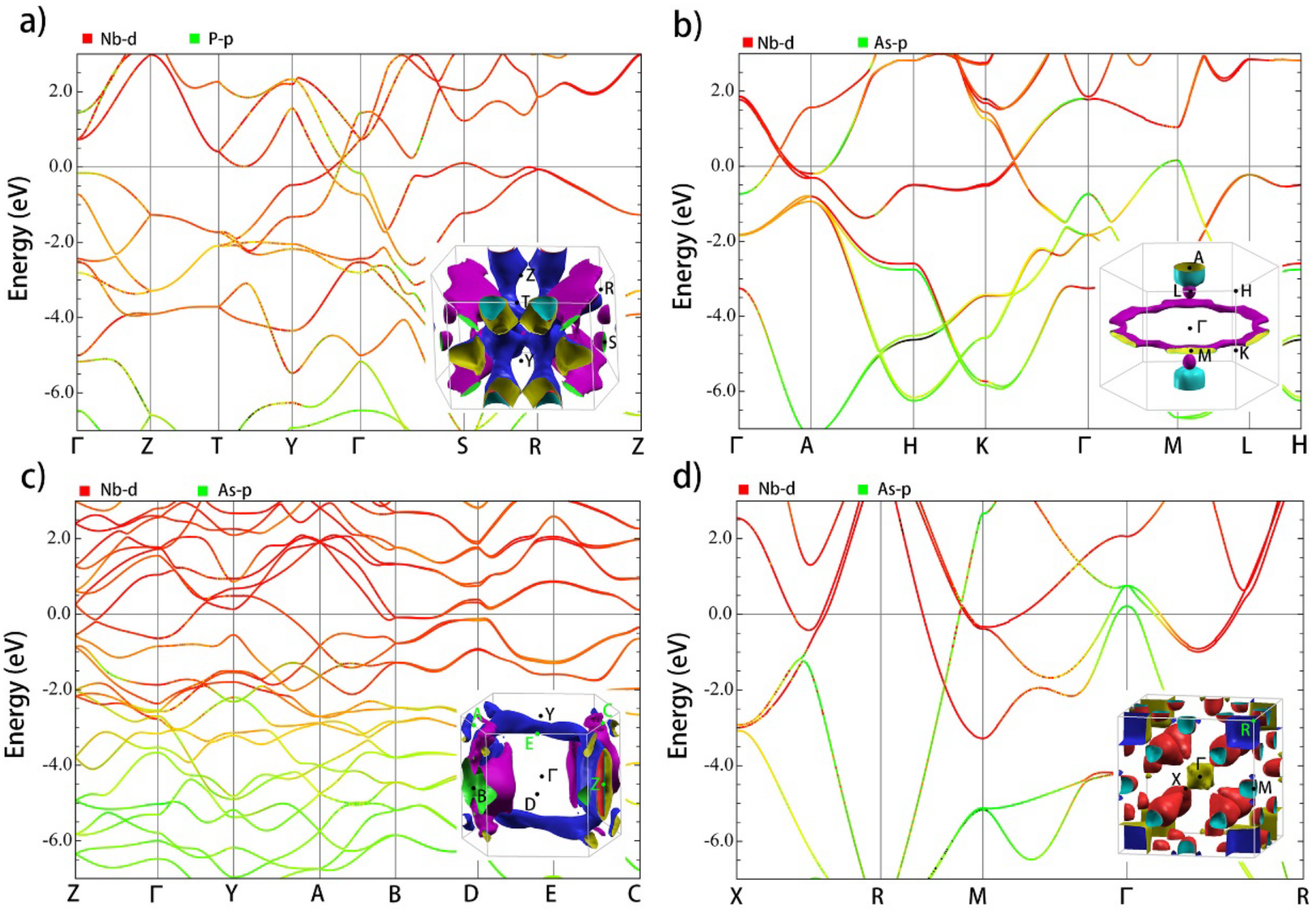

Fig 3 Band structures and Fermi surfaces of NbP (a) and NbAs (b-d) high pressure phases. The contribution of Nb atom's d orbitals and P(As) atom's p orbitals are marked with red and green in band structures, respectively. The Fermi pockets of NbP and NbAs high pressure phases are mainly contributed from Nb atom's d orbitals. However, the pockets in *Pm*-*3m*-NbAs around Γ point are contributed from As atom's p orbitals.

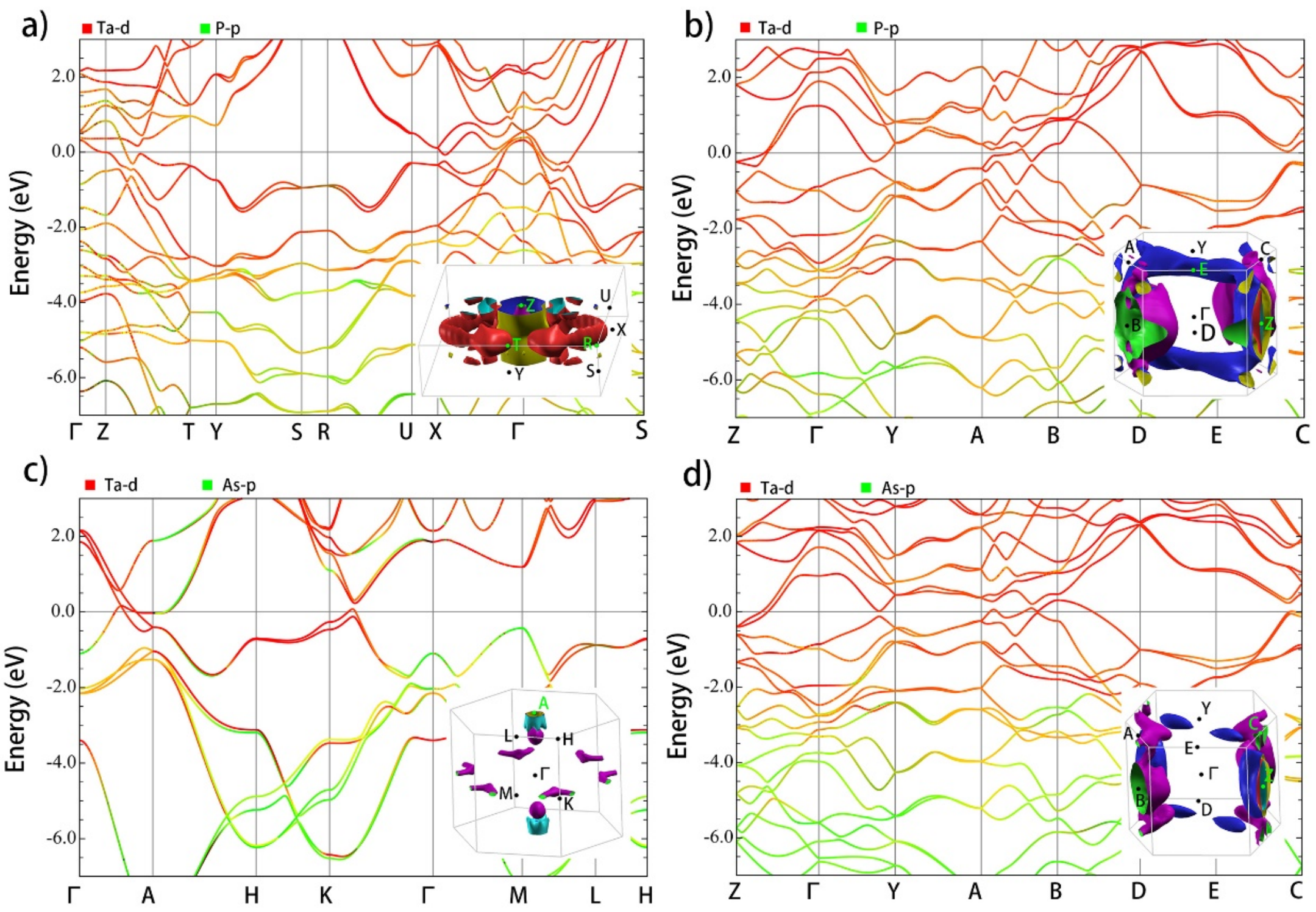

Fig 4 Band structures and Fermi surfaces of TaP (a,b) and TaAs (c,d) high pressure phases. The contribution of Ta atom's d orbitals and P(As) atom's p orbitals are marked with red and green in band structures, respectively. The bands near the Fermi level are contributed by Ta atom's d orbital.

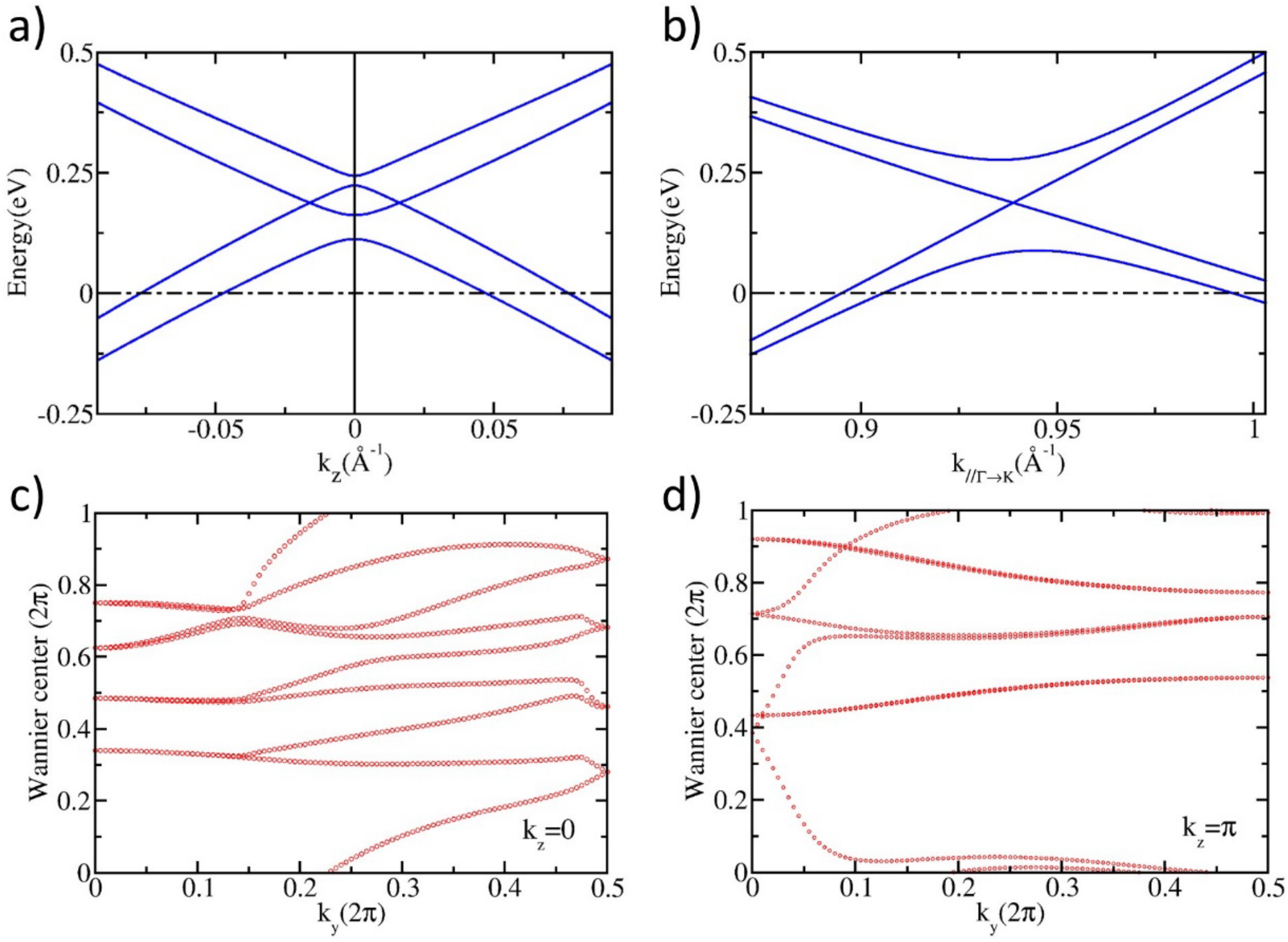

Fig 5 The detailed band dispersions and $Z_2$ topological invariants calculations for the high pressure *P-6m2* NbAs. (a,b) Band structure around the Weyl point along the path parallel with $k_z$ direction a) and $\Gamma - K$ high symmetry line b). The Fermi energy is set to 0eV. (c,d) $Z_2$ topological invariants calculated by Wilson loop method along the $k_y$ axis in $k_z$=0 (c) and $k_z=\pi$ (d) planes, both of them are non-trivial.

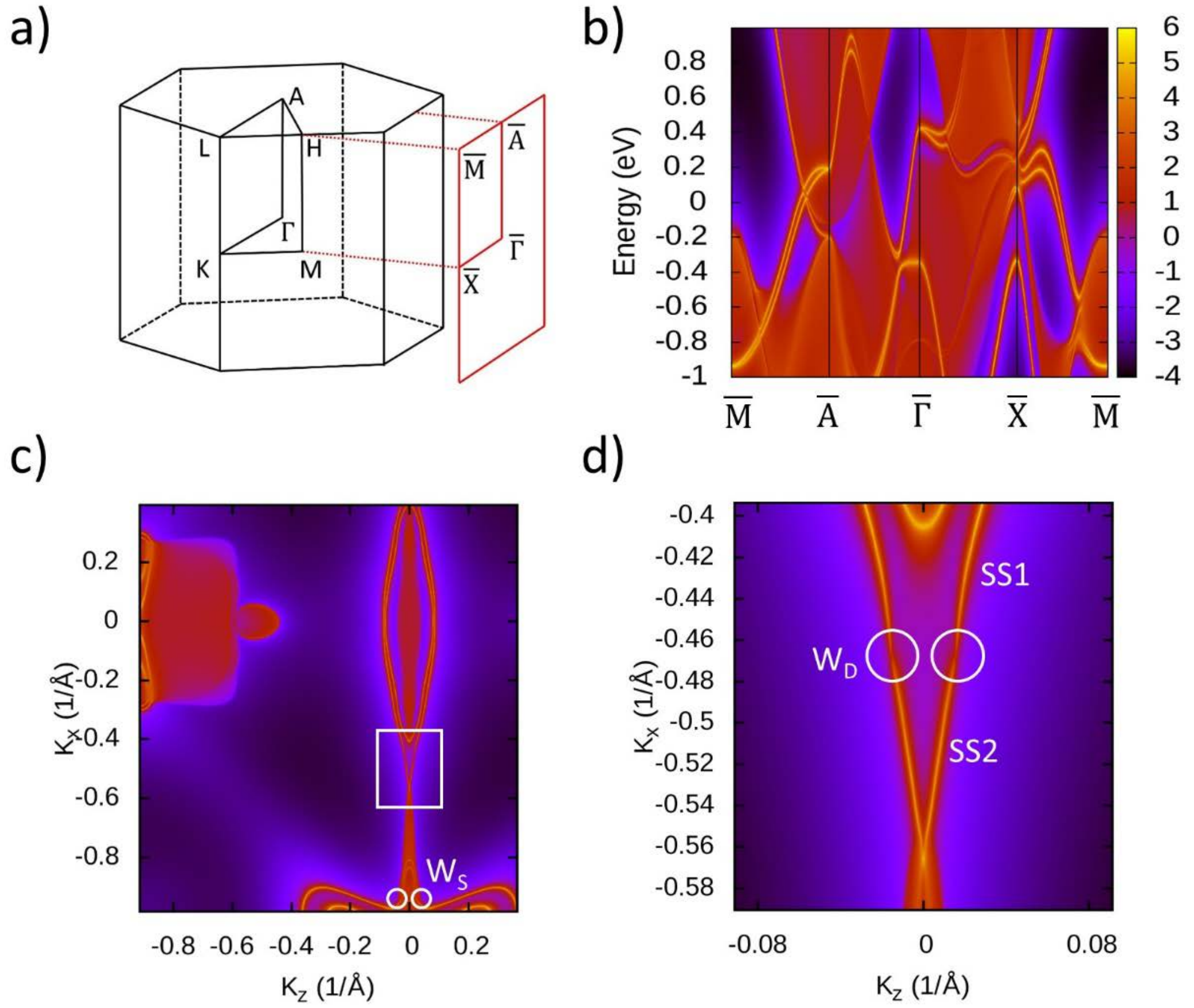

Fig 6 The surface states of the high pressure *P-6m2* NbAs. a) The first BZ of *P-6m2*-NbAs and (100) projected plane of the BZ b) Projected surface density of state of (100) plane with As termination. c,d) Fermi surfaces (c) with the chemical potential at energy of Weyl points (177meV), and details in the white frame in (c) are enlarged and shown in (d). (d) The Weyl points are marked with white circle, the Fermi arcs are marked with SS1 and SS2. The projection points of single Weyl point and two Weyl points are marked by $W_S$ and $W_D$, respectively.

**Table I,** Structural information of calculated and experimental structures of TaAs family. The structural information includes the space group, cell parameter, and Wyckoff positions.

| | Pressure (GPa) | Space-group | a(Å) | b(Å) | c(Å) | $\alpha$(°) | $\beta$(°) | $\gamma$(°) | Wyckoff position |
|---|---|---|---|---|---|---|---|---|---|
| NbP | 0 | $I4_1md$ | 3.437 | 3.437 | 11.656 | 90 | 90 | 90 | Nb:4a(0.500, 0.500, 0.833)<br>P:4a(0.500, 0.500, 0.250) |
| | exp[61] | $I4_1md$ | 3.332 | 3.332 | 11.371 | 90 | 90 | 90 | |
| | 70 | Cmcm | 2.925 | 8.411 | 3.913 | 90 | 90 | 90 | Nb:4c(0.000, 0.861, 0.250)<br>P:4c(0.000, 0.585, 0.250) |
| TaP | 0 | $I4_1md$ | 3.338 | 3.338 | 11.401 | 90 | 90 | 90 | Ta:4a(0.500, 0.500, 0.833)<br>P:4a(0.500, 0.500, 0.250) |
| | exp[62] | $I4_1amd$ | 3.33 | 3.33 | 11.39 | 90 | 90 | 90 | |
| | 70 | Pmmn | 3.194 | 3.028 | 10.338 | 90 | 90 | 90 | Ta1:2a(0.000, 0.000, 0.083)<br>Ta2:2b(0.000, 0.500, 0.333)<br>P1:2a(0.500, 0.500, 0.167)<br>P2:2b(0.500, 0.000, 0.433) |
| | 90 | $P2_1/c$ | 4.455 | 4.499 | 7.964 | 90 | 143.7 | 90 | Ta:4e(0.897, 0.855, 0.104)<br>P:4e(0.351, 0.857, 0.582) |
| NbAs | 0 | $I4_1md$ | 3.452 | 3.452 | 11.68 | 90 | 90 | 90 | Nb:4a(0.000, 0.500, 0.917)<br>As:4a(0.000, 0.500, 0.333) |
| | exp[63] | $I4_1md$ | 3.452 | 3.452 | 11.679 | 90 | 90 | 90 | |
| | 35 | P-6m2 | 3.186 | 3.186 | 3.4279 | 90 | 90 | 120 | Nb:1f(0.667, 0.333, 0.500)<br>As:1c(0.333, 0.667, 0.000) |
| | 40 | $P2_1/c$ | 4.766 | 4.78 | 8.502 | 90 | 143.6 | 90 | Nb:4e(0.601, 0.646, 0.895)<br>As:4e(0.140, 0.852, 0.916) |
| | 70 | Pm-3m | 2.963 | 2.963 | 2.963 | 90 | 90 | 90 | Nb:1a(0.000, 0.000, 0.000)<br>As:1b(0.500, 0.500, 0.500) |
| TaAs | 0 | $I4_1md$ | 3.465 | 3.465 | 11.726 | 90 | 90 | 90 | Ta:4a(0.500, 0.000, 0.250)<br>As:4a(0.500, 0.000, 0.667) |
| | exp[64] | $I4_1md$ | 3.437 | 3.437 | 11.644 | 90 | 90 | 90 | |
| | 20 | P-6m2 | 3.27 | 3.27 | 3.448 | 90 | 90 | 120 | Ta:1f(0.667, 0.333, 0.500)<br>As:1c(0.333, 0.667, 0.000) |
| | 40 | $P2_1/c$ | 5.551 | 4.753 | 5.765 | 90 | 130.3 | 90 | Ta:4e(0.316, 0.856, 0.207)<br>As:4e(0.809, 0.854, 0.226) |